\documentclass[%
reprint,
superscriptaddress,
aps,
prb,
]{revtex4-2}

\usepackage{graphicx}
\usepackage{dcolumn}
\usepackage{bm}
\usepackage{physics}
\usepackage{amssymb}
\usepackage[caption=false]{subfig}
\usepackage{color}
\usepackage{ulem}
\usepackage{float}

\usepackage[squaren]{SIunits}
\usepackage{hyperref}
\hypersetup{colorlinks=true,linkcolor=blue,citecolor=blue,urlcolor=blue}

\begin{document}


\title{Anderson mobility edge as a percolation transition}

\author{Marcel Filoche}
\affiliation{Institut Langevin, ESPCI, CNRS, Universit\'e PSL, 75005 Paris, France}
\email{marcel.filoche@espci.psl.eu}

\author{Pierre Pelletier}
\affiliation{
Laboratoire de Physique de la Mati\`ere Condens\'ee, Ecole Polytechnique, CNRS, Institut Polytechnique de Paris, 91120 Palaiseau, France
}

\author{Dominique Delande}
\affiliation{
Laboratoire Kastler Brossel, Sorbonne Universit\'e, CNRS, ENS-PSL Research University,
Coll\`ege de France, 4 Place Jussieu, 75005 Paris, France
}

\author{Svitlana Mayboroda}
\affiliation{
School of Mathematics, University of Minnesota, Minneapolis, Minnesota 55455, USA
}

\date{\today}

\begin{abstract}
The location of the mobility edge is a long standing problem in Anderson localization. In this paper, we show that the effective confining potential introduced in the localization landscape (LL) theory predicts the onset of delocalization in 3D tight-binding models, in a large part of the energy-disorder diagram. Near the edge of the spectrum, the eigenstates are confined inside the basins of the LL-based potential. The delocalization transition corresponds to the progressive merging of these basins resulting in the percolation of this classically-allowed region throughout the system. This approach, shown to be valid both in the cases of uniform and binary disorders despite their very different phase diagrams, allows us to reinterpret the Anderson transition in the tight-binding model: the mobility edge appears to be composed of two parts, one being understood as a percolation transition.
\end{abstract}

\maketitle



In three dimensions and above, tight-binding models with on-site diagonal disorder (à la Anderson) are characterized by the existence of a transition between exponentially localized states and delocalized states~\citep{Anderson1958, Abrahams1979, Lagendijk2009} called the \emph{mobility edge} (ME). In the energy-disorder phase diagram this ME is materialized by a line where a second-order phase transition occurs: on the localized side, the localization length diverges near the line while on the delocalized side, the system conductivity vanishes~\cite{Grussbach1995}.

Computing the specific location of the mobility edge and the critical exponents associated to this transition is a difficult task, mainly because it takes place in the strong disorder regime where perturbative approaches fail. Non-perturbative approaches have been tried to predict the position of the ME, one of the more advanced being the self-consistent theory of localization~\cite{Kroha1990, Vollhardt1992, Wolfle2010}. It has been used for the specific case of cold atoms in a disordered optical potential in~\cite{Kuhn2007, Yedjour2010, Piraud2013}. However, the results depend on the approximations, and only a semi-quantitative agreement between the experimental results~\cite{Jendrzejewski2012, Semeghini2015} and the numerical calculations~\cite{Delande2014, Pasek2017} has been obtained for the position of the mobility edge. Moreover, this approach brings very few information about the detailed spatial structure of the wave functions and the mechanism at work at the localization/delocalization transition.

In this work we show that a proper renormalization of the disorder, performed by the localization landscape (LL) theory, predicts the onset of delocalization in a large part of the phase diagram for the tight-binding model in~3D and brings new insights on the nature of the transition in the spatial realm. To that end, we first introduce the Hamiltonian~$\hat{H}_0$ on $\mathbb{Z}^3$:
\begin{equation}\label{eq:sym_Hamiltonian}
\left(\hat{H}_0 \psi\right)_n = -\sum_{m \in \ev{n}} \psi_m + W \nu_n \, \psi_n  \,,
\end{equation}
where $\ev{n}$ stands for the set of nearest neighbors of site~$n$, $W$ is a non negative quantity (the disorder strength), and $\nu_n$ is an i.i.d. random variable of uniform (or box) law in the interval~$[-1/2,1/2]$. When a finite size lattice such as a cube of size $N \times N \times N$ is studied, see below), we use periodic boundary conditions in the 3~directions of space.

The LL introduced in~\cite{Filoche2012} is defined as the solution to
\begin{equation}\label{eq:landscape}
\hat{H} u = 1 \,,
\end{equation}
where $\hat{H}$ is the Hamiltonian and the right-hand side is a vector with all entries equal to~1 (it is the discrete counterpart to the constant function~$1$ in the continuous setting). Since the Hamiltonian $\hat{H}_0$ has a statistically symmetric spectrum in $[-6 -W/2; 6 + W/2]$, we need to perform an energy shift by the quantity $6 + W/2$ in order to obtain a positive~LL, this quantity being also the minimal and optimal one~\cite[Sec. II.D]{Filoche2017}. The shifted Hamiltonian reads:
\begin{equation}\label{eq:asym_Hamiltonian}
(\hat{H} \psi)_n = -\sum_{m \in \ev{n}} (\psi_m -\psi_n) + V_n \, \psi_n \,,
\end{equation}
with $V_n = W \left( \nu_n + 0.5 \right)$. Its spectrum lies in $[0, W+12]$. 

One of the main results of the LL theory is that the quantity~$1/u$ defines an effective potential that possesses several remarkable properties for understanding localization: (i) its basins predict the region of localization of the low energy states, (ii) it can be used to build a very good approximation to the integrated density of states (the number of states below a given energy)~\cite{Arnold2016, Arnold2019b, David2021}, and (iii) its structure governs the long range behavior of the localized wave functions~\cite{Arnold2016, Arnold2019a}. One can in fact envision this potential as the result of a spatial renormalization of the original potential~$V$ at a local self-adapted scale.

To support this statement, one simply has to conjugate the Schr\"odinger equation with the LL, which amounts to introducing the auxiliary field $\varphi$ such that $\psi = u \varphi$. This allows us to rewrite the action of $\hat{H}$ on $\psi$ in the following way (see Appendix for the detailed derivation):
\begin{align}\label{eq:effective_1}
(\hat{H}\psi)_n &= -\sum_{m \in \ev{n}} (\psi_m - \psi_n) + V_n \psi_n \nonumber\\
&= -\sum_{m \in \ev{n}} u_m ( \varphi_m - \varphi_n)  ~+~ \varphi_n \,.
\end{align}

Consequently, the average energy of any wave function~$\psi$ can be also rewritten in terms of $\varphi$~\cite{Wang2021AnnHenriPoincare, Arnold2022CommMathPhys}:
\begin{align}\label{eq:effective_2}
\expval{\hat{H}}{\psi} &= \sum_{(m,n)} u_n u_m (\varphi_m - \varphi_n)^2 + \sum_n u_n \varphi_n^2 \nonumber\\
& = \sum_{(m,n)} u_n u_m \left(\frac{\psi_n}{u_n} - \frac{\psi_m}{u_m} \right)^2 + \sum_n \frac{1}{u_n} \, \psi_n^2  \,,
\end{align}
where the first sum runs on all edges $(m,n)$ of the lattice (connecting nearest neighbors), each edge being counted only once, while the second sum is the average value of $1/u$ in the state~$\psi$. Both sums being non negative, it follows that for all wave functions~$\psi$, $\expval{\hat{H}}{\psi} \ge \expval{\widehat{1/u}}{\psi}$. In other words, this second term can be seen as an effective potential energy governing the structure of the state. In fact, Eq.~(\ref{eq:effective_2}) is the discrete counterpart of identity~(6) in~\citep{Arnold2016} found in the continuous setting.

\begin{figure}[ht!]
\includegraphics[width = 0.23\textwidth]{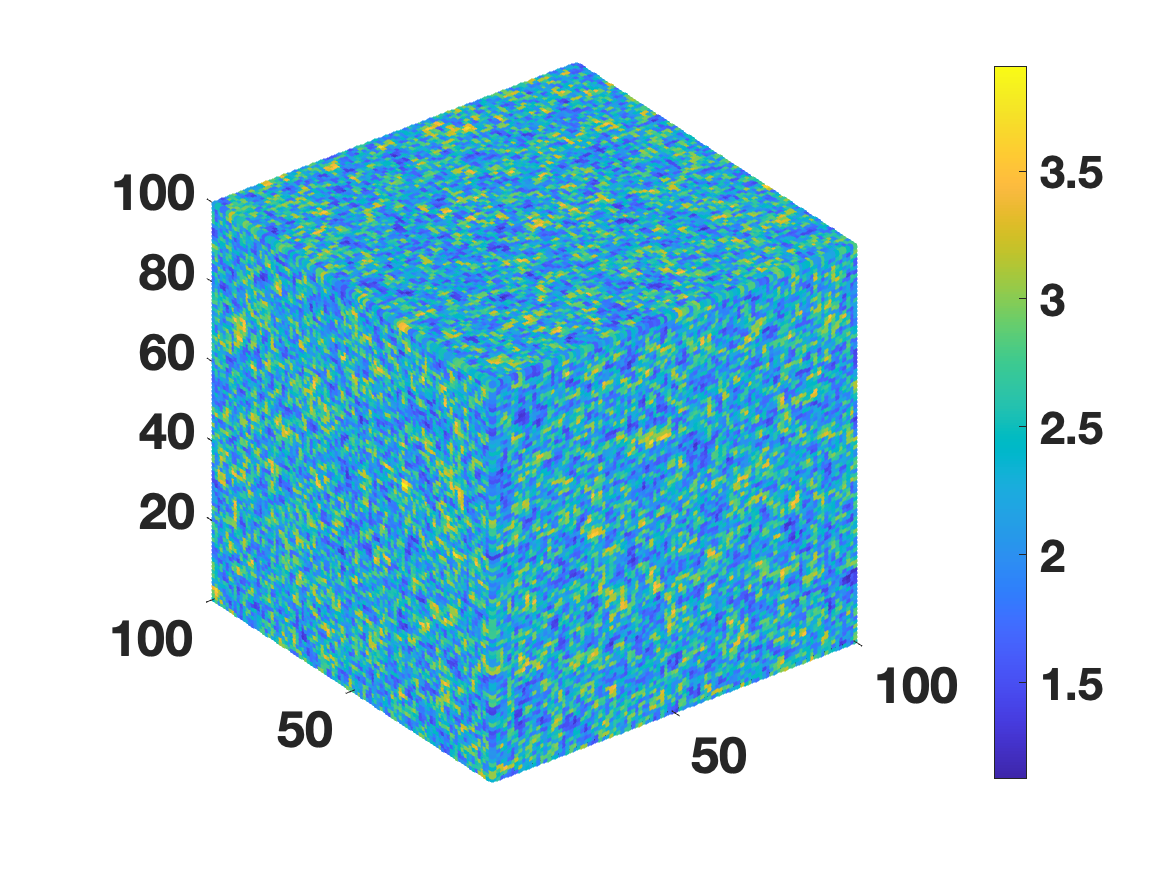}
\includegraphics[width = 0.23\textwidth]{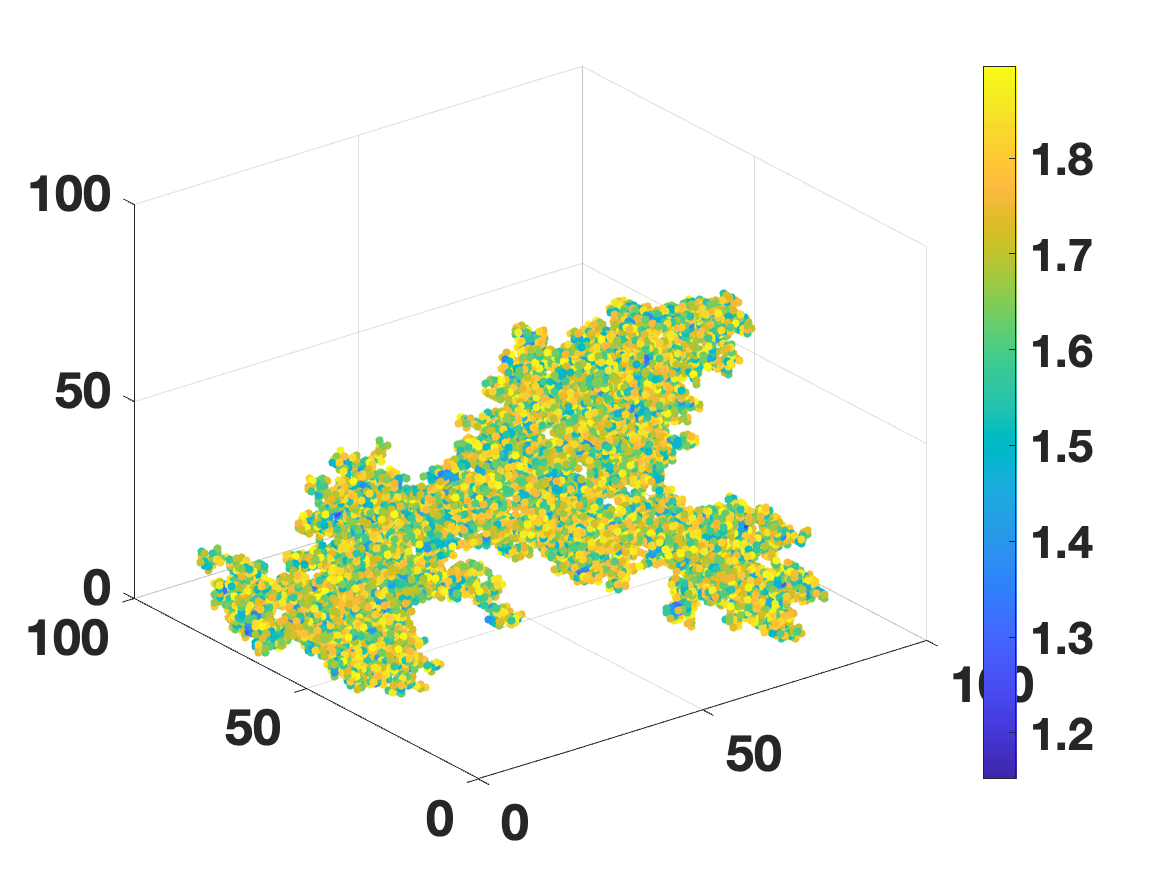}
(a) \hskip 4 cm (b)
\caption{(a)~3D map of the effective potential $1/u$, for a random potential with uniform law and disorder strength $W=5$, computed on a cube of lateral size $N=100$. (b)~Percolating cluster at energy $E_c$ corresponding to the realization on the left. This cluster is the percolating subset of the domain $\Omega_E = \left\{ n \in \Omega, 1/u_n \le E \right\}$, for the smallest energy $E$ for which this domain contains a path joining any two opposite faces of the cube. 
}
\label{fig:cluster_Ec}
\end{figure}

The quantity $1/u$ therefore plays the role of a classical potential, while being much more regular than~$V$ because it is the solution to a proper second-order PDE. The core hypothesis of this work is that the eigenfunctions are confined by $1/u$. As a consequence, in order for an eigenfunction~$\psi$ of $\hat{H}$ of eigenvalue~$E$ to delocalize, the classically allowed region of the LL effective potential, defined as $\Omega_E = \left\{ n \in \Omega, 1/u_n \le E \right\}$, has to percolate through the entire domain~$\Omega$. When $\Omega$ is a cube, we thus define the \emph{LL~percolation threshold}~$E_c$ as the smallest energy for which $\Omega_E$ contains a path between neighboring sites connecting any two opposite faces. Figure~\ref{fig:cluster_Ec}a displays a 3D~representation of the computed values of $1/u$ for one realization of a uniform disorder, while Fig.~\ref{fig:cluster_Ec}b shows the percolation cluster of the subset $\Omega_{E_c}$ for the same realization of the disorder.

For each value of the disorder strength~$W$, we compute the average~$\bar{E}_c$ over 20~realizations of the disorder for a system of lateral size~$N=50$. In order to compare these values with values of the ME found in the literature, one first has to subtract the spectral shift $(6+W/2)$ between the symmetric Hamiltonian of Eq.~\eqref{eq:sym_Hamiltonian} and the asymmetric Hamiltonian of Eq.~\eqref{eq:asym_Hamiltonian} used to compute the effective potential~$1/u$. This gives a negative value of the percolation threshold, which corresponds to the negative part of the spectrum of $\hat{H}_0$. However, we can take advantage of the statistical symmetry of its spectrum (see Appendix) and perform the comparison between the ME and the percolation threshold of~$1/u$ on the positive energy side of the energy-disorder diagram by defining the percolation threshold $E_c^*$ as:
\begin{equation}\label{eq:Ec*}
E_c^*(W) = 6 + W/2 - E_c(W) .
\end{equation}

\begin{figure}[ht!]
\includegraphics[width = 0.45\textwidth]{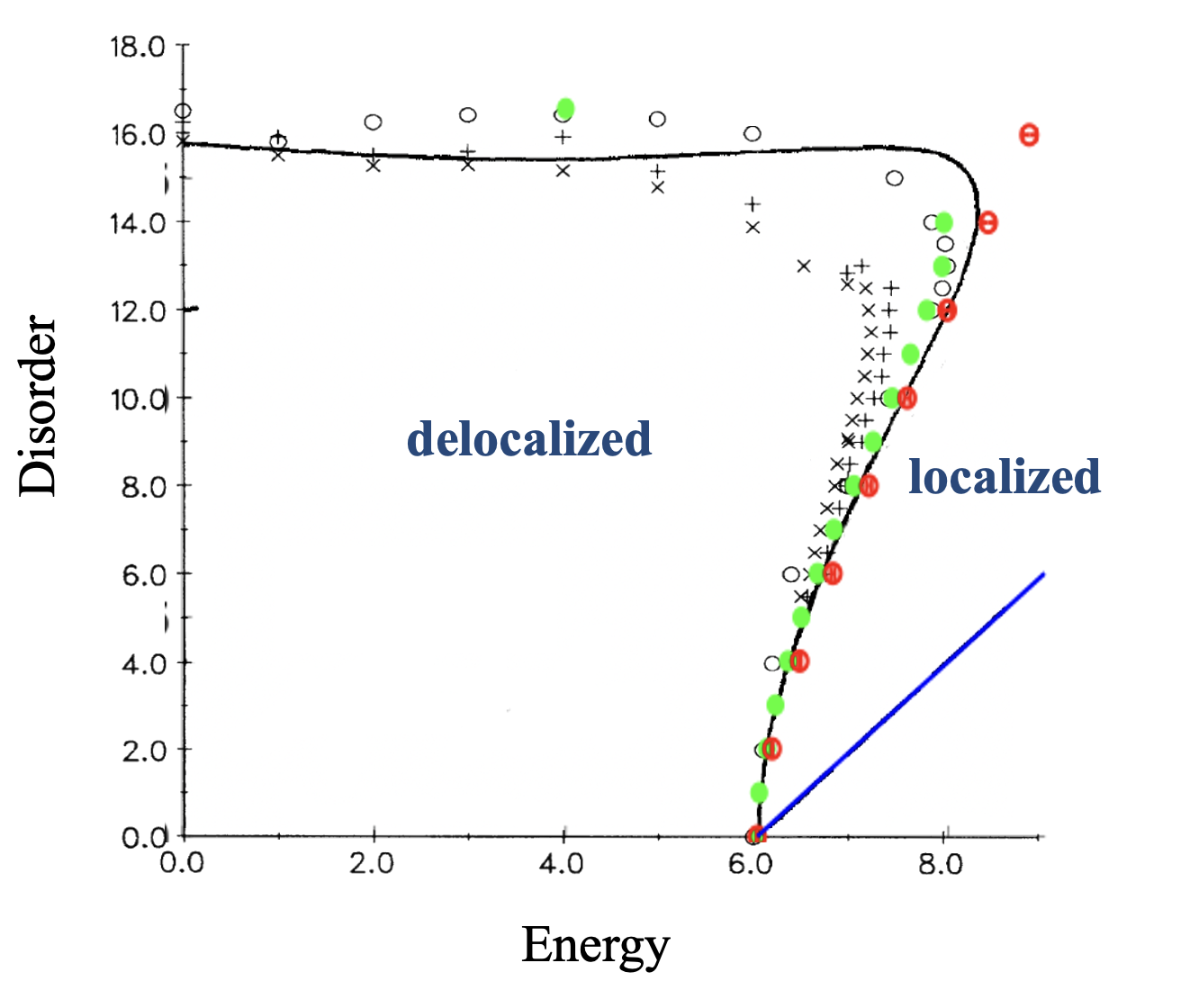}
\caption{Phase diagram of localization in the (energy, disorder) plane, for a uniform disorder. Due to the $E \rightarrow -E$ symmetry of the problem, only the positive part $(E\ge 0$) of the diagram is represented. The straight blue line to the right represents the edge of the spectrum. The black crosses and circles correspond to simulation data from Grussbach and Schreiber~\cite{Grussbach1995}, while the red circles show the percolation threshold of the LL~effective potential (on a domain of size $N=50$ with 20~samples for each point) and the green filled circles show our computations of the ME using the TMM. The continuous black line is the prediction of the self-consistent theory~\cite{Kroha1990}.
}
\label{fig:uniform_ME}
\end{figure}

\begin{figure}[ht!]
\includegraphics[width = 0.23\textwidth]{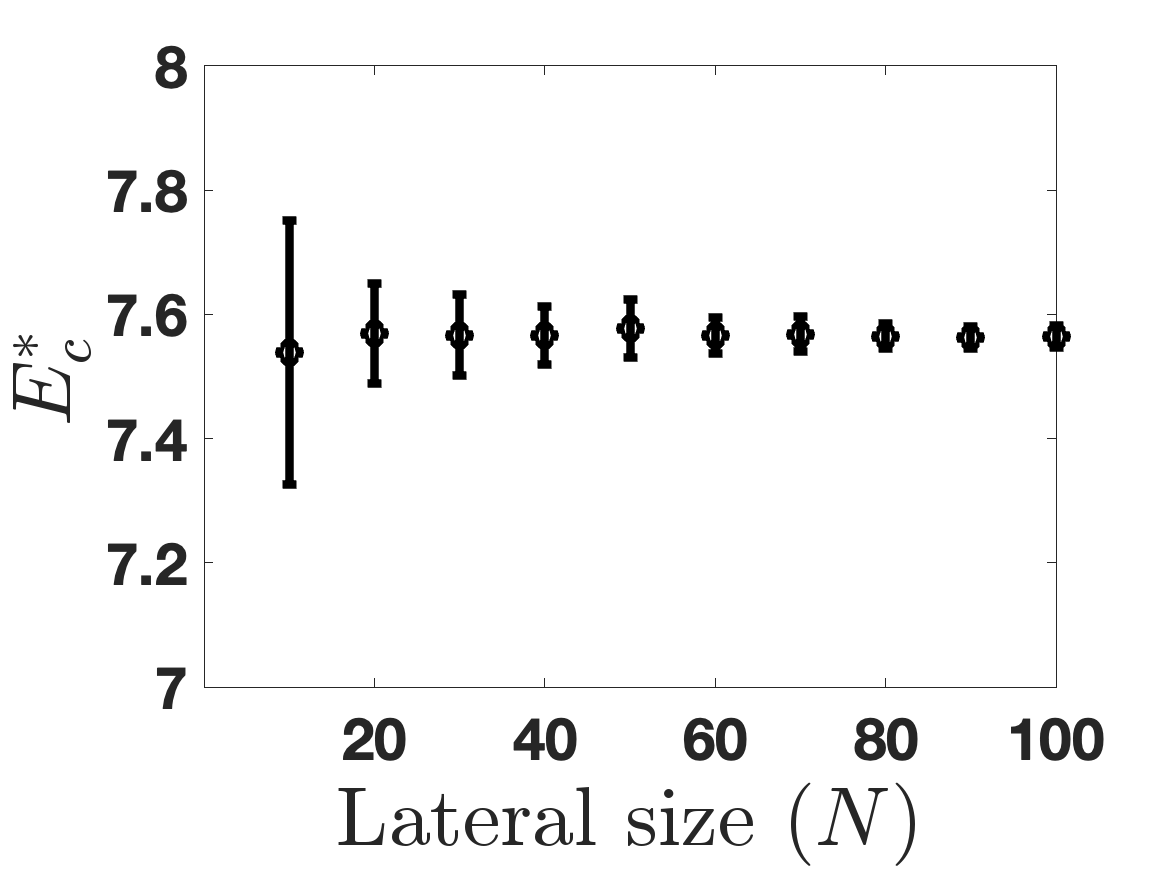}
\includegraphics[width = 0.23\textwidth]{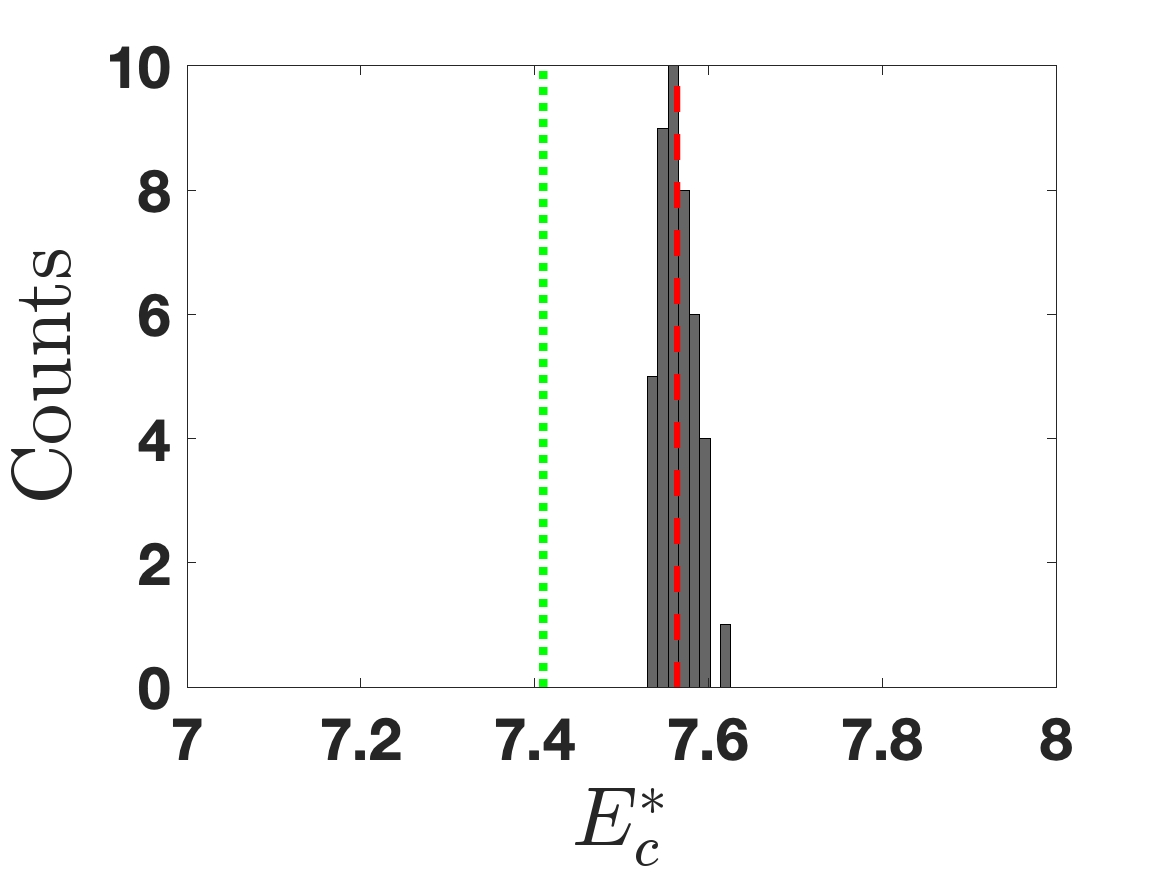}
(a) \hskip 4 cm (b)
\caption{(a)~Average and standard deviation of the LL~percolation threshold $E_c^*$, as defined in Eq.~\eqref{eq:Ec*}, for various system sizes (from $N=10$ to $N=100$), computed on~50 realizations of the disorder (uniform Anderson model, disorder strength $W=10$). (b)~Histogram of the values of $E_c^*$ at $N=50$ for 50~different realizations (uniform Anderson model, disorder strength $W=10$). The average value $E_c^*$ is represented by a red dashed line. One observe a standard deviation smaller than~0.1. The value of the ME computed by the TMM is represented by the vertical dashed green line.
}
\label{fig:histogram}
\end{figure}

We also performed TMM computations in order to cross-check the results from the literature with new simulations on larger samples and to ensure consistency of our results. Figure~\ref{fig:uniform_ME} displays the energy-disorder diagram in which the computed average values of $E_c^*$ (red circles) are superimposed with our computed values of the ME using the transfer matrix method (TMM, green filled circles), as well as values from~\cite{Bulka1987, Grussbach1995} obtained either by the TMM (black circles) or by multifractal analysis (black crosses, '$+$' and '$\times$'). The blue line is the edge of the spectrum. One can clearly see a very good agreement between the LL~percolation threshold and the ME, up to a disorder strength $\sim 14$, a value close to the critical disorder $\sim16.5$ known in the literature~\cite{Rodriguez2010PhysRevLett, Slevin2014NewJPhys,  Slevin2018JPhysSocJpn}. For larger disorder strength, the values of the percolation threshold continue to grow towards the right of the graph while the mobility edge line bends back to become almost horizontal. The standard deviation of $E_c^*$ for a size $N=50$ and a given disorder strength~$W$ are about \unit{5.10^{-2}}, as shown in the graph of Fig.~\ref{fig:histogram}a and the histogram of Fig.~\ref{fig:histogram}b. These (horizontal) error bars are almost invisible in~Fig.~\ref{fig:uniform_ME}. The ME computed by the TMM for the same value $W=10$ is represented by a vertical dashed green line in Fig.~\ref{fig:histogram}b. The percolation threshold of $1/u$ appears slightly larger (hence closer to the band edge) than the ME, again, up to disorder strength $\sim$14.

\begin{figure}[ht!]
\includegraphics[width = 0.45\textwidth]{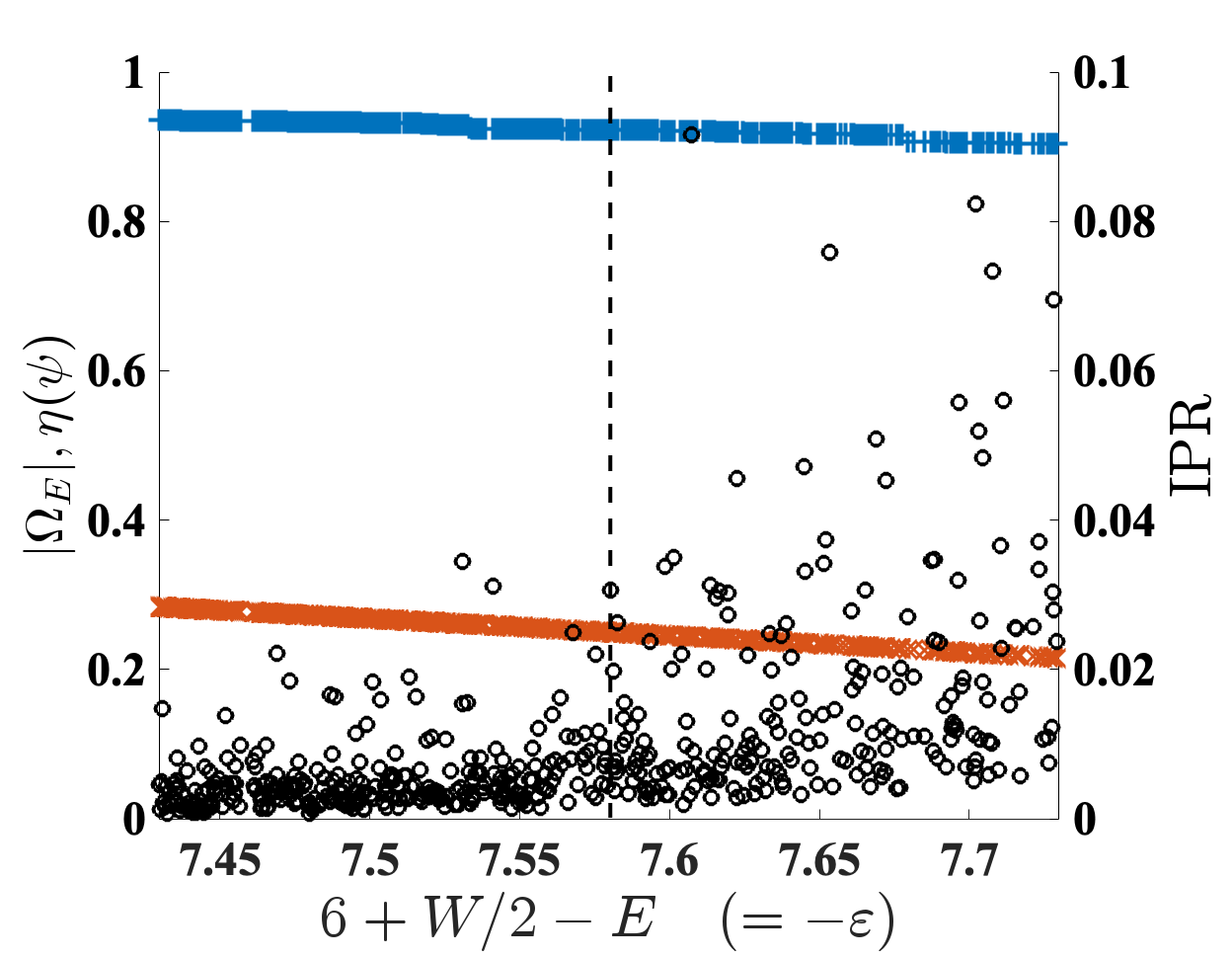}
\caption{Plots of the relative size of the set~$\Omega_E$ (red crosses '$\times$'), and the values of $\textrm{IPR}(\psi)$ (black circles) and $\eta(\psi)$ (blue crosses '+'), as defined in Eq.~\eqref{eq:PR_eta}, for 500~eigenvalues around the LL percolation threshold at $W=10$ (represented by the vertical dashed line). Although the set $\Omega_E$ occupies only about 25\% of the total domain at the LL percolation threshold, one sees from that the eigenstates are almost entirely confined inside this set ($\eta(\psi) \gtrsim 0.9$). The sharp increase of $\textrm{IPR}(\psi)$ (circles) across the LL percolation threshold confirms that strong localization of the eigenstates occurs in the vicinity of this threshold ($\textrm{IPR}(\psi)\sim 1/10$ means that $\psi$ is localized on about 10~sites). The computation is performed about a system of lateral size $N=50$.
}
\label{fig:PR}
\end{figure}

To validate our confinement hypothesis by the effective potential, we have to show that the eigenstates of $\hat{H}_0$ with (negative) energy $\varepsilon$ delocalize across the LL~percolation threshold while nevertheless remaining confined to the set~$\Omega_E$, where $E=\varepsilon + 6 + W/2$. To that end, we compute for a given disorder strength~$W$ and for each eigenvalue in a small interval around $-E_c^*$ the inverse participation ratio (IPR) as well as the fraction~$\eta(\psi)$ of the corresponding eigenfunction lying in~$\Omega_E$, defined as:
\begin{equation}\label{eq:PR_eta}
\textrm{IPR}(\psi) = \sum_n \psi_n^4 \quad, \quad \eta(\psi) = \sum_{n \in \Omega_E} \psi_n^2 \,
\end{equation}
where $\psi$ is assumed to be $L^2$-normalized (i.e., $\sum_n \psi_n^2 = 1$). If $\psi$ is constant on a set of size $M$ and vanishes elsewhere, then $\textrm{IPR}(\psi)=1/M$. In other words, $\textrm{IPR}(\psi)$ assesses the typical number of sites over which $\psi$ is supported. Regarding~$\eta(\psi)$, a value~${\sim}1$ means that the eigenstate~$\psi$ is almost entirely confined inside the set~$\Omega_E$. Figure~\ref{fig:PR} displays both quantities plotted against the energy of the eigenstate, for 500~energies around the LL percolation threshold ($W=10$, $N=50$). On the same graph is plotted the size of the set~$\Omega_E$ (red crosses '$\times$'). One can see that the set~$\Omega_E$ occupies about 25\% of the total volume at the percolation threshold. Yet, $\eta(\psi)$ is larger than 0.9 for all eigenstates (blues crosses '+'), which means that all eigenstates are mostly supported on~$\Omega_E$. Finally, the steep increase of the values of IPR at $E\approx E_c$ confirms that the localization-delocalization transition occurs very close to the LL percolation threshold.

\begin{figure}[ht!]
\includegraphics[width = 0.45\textwidth]{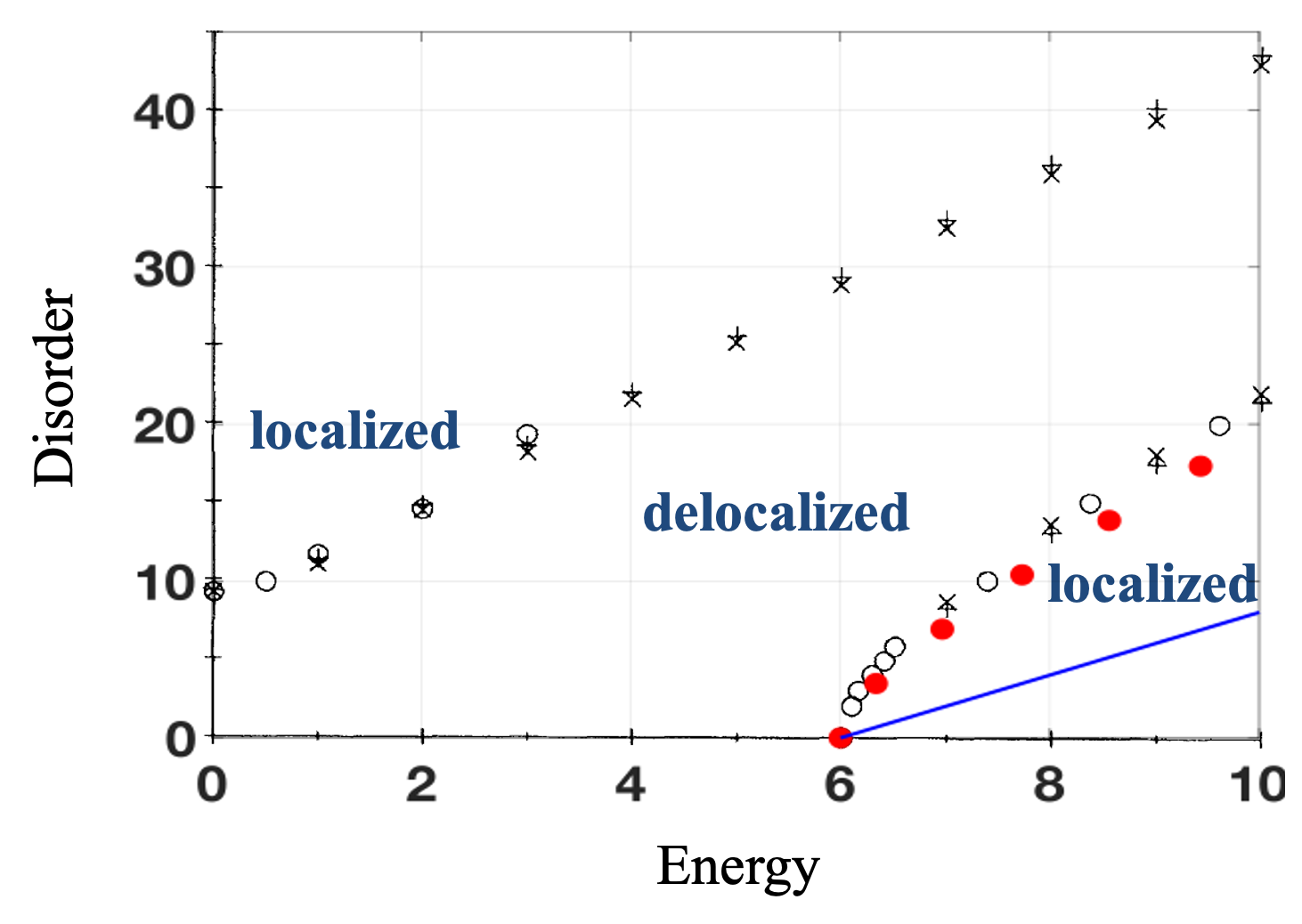}
\caption{Phase diagram of localization in the (energy, disorder) plane, for a binary disorder. The solid blue line represents the edge of the spectrum. The black crosses and circles correspond to simulation data from Grussbach and Schreiber~\cite{Grussbach1995}, while the red dots show the percolation threshold of the LL~effective potential (computed on a domain of size $N=50$ for a disorder strength that is $\sqrt{3}$~times larger to remain consistent with the convention of~\cite{Grussbach1995}).
}
\label{fig:binary_ME}
\end{figure}

The same approach is applied to the study of the Anderson binary model in which the random law $\nu_n$ in Eq.~\eqref{eq:sym_Hamiltonian} is a Boolean law on~$\{-1/2, 1/2\}$. In this case, the localization pattern in the energy-disorder diagram exhibits a very different behavior from the Anderson uniform model (see~Fig.~\ref{fig:binary_ME}). Instead of being confined to a bounded region of the diagram, the delocalized phase is present at any strength of the disorder~\cite{Grussbach1995}. Consequently, the ME exhibits three independent branches, two symmetric near the edges of the spectrum (near $-6-W/2$ and $6+W/2$), and one branch corresponding to the critical disorder around $W=10$ at $\varepsilon=0$ (only the positive energy region is represented here).

The LL percolation threshold is computed for several values of the disorder strength on a cube of lateral size~$N=50$ (averaged over 20~samples), and the values are superimposed on the phase diagram of Fig.~\ref{fig:binary_ME}. In this case also, we observe a striking similarity between the LL~percolation threshold and the branch of the ME close to the edge of the spectrum. Given the universality of the LL procedure, we believe this prediction of the lateral branch to be very general and not related to a specific type of disorder. In particular, it should also predict the mobility edge in lattices or networks~\cite{Razo-Lopez2023} where the TMM is much more difficult to compute.

These results show that the position of the ME is closely linked to the properties of the set~$\Omega_E$ in the vicinity of its percolation threshold. One has to stress that, contrary to the original potential~$V$ which is an i.i.d. random variable, $1/u$ is not: it contains spatial correlations which make its statistical properties and in particular its percolation threshold more difficult to compute.

\begin{figure}[ht!]
\includegraphics[width = 0.40\textwidth]{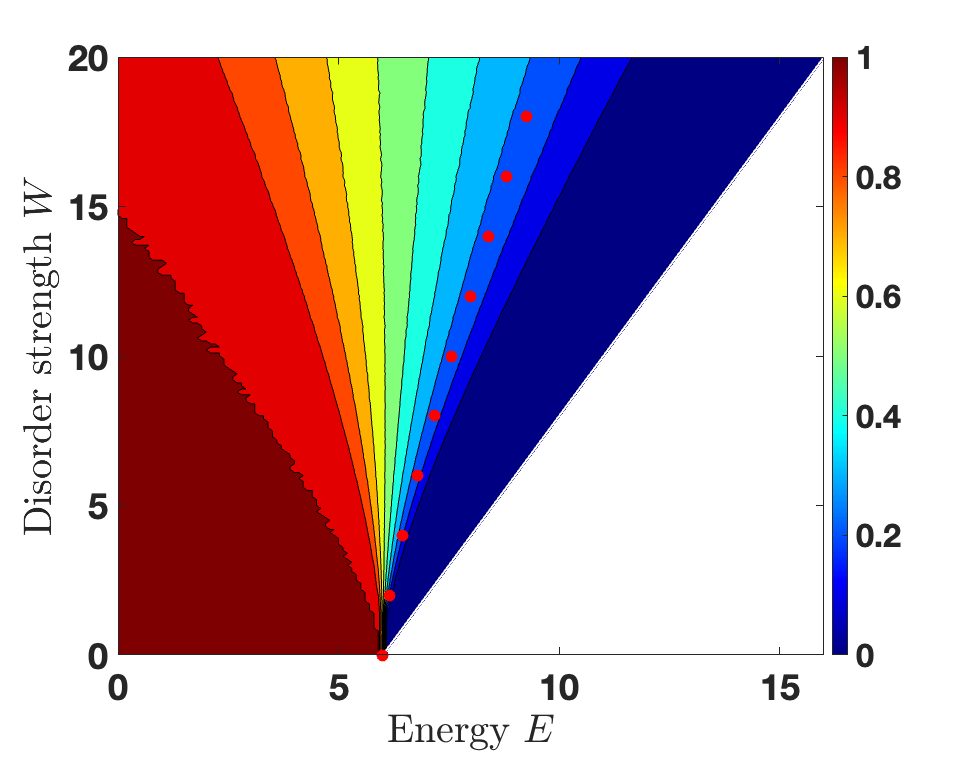}
\caption{Contour plot of the relative size of the set~$\Omega_E$ in the energy-disorder diagram, computed on a domain of lateral size~$N=40$. The red dots represent the LL~percolation threshold displayed in Fig.~\ref{fig:uniform_ME}. One can see that they correspond to a constant relative size of $\Omega_E$, about 25\% of the entire domain. On the vertical axis $E=0$, the relative size 1 is attained very close to the known value of the ME on this axis.
}
\label{fig:Size_OE}
\end{figure}

To compare the percolation properties of $\Omega_E$ with classical percolation, we compute (for the Anderson uniform disorder) the size of $\Omega_E$ relative to the total size of the domain. Figure~\ref{fig:Size_OE} displays a contour plot of this relative size in the energy-disorder diagram, computed on a domain of lateral size~$N=40$ for the Anderson uniform disorder. The values of the LL percolation threshold are superimposed on this plot (red dots). They are found to occur always at the same fraction of the total size of the domain, about 25\%, independently of the strength of the disorder $W$, a feature consistent with the universal properties of percolation. Indeed, if the values of $1/u$ at each site were independent random variables drawn uniformly in an interval, then the percolation of $\Omega_E$ would occur at a value $E$ such that about 31.2\% of the total number of sites would satisfy $1/u_n \le E$, a fraction corresponding to the site percolation threshold of the cubic lattic $p_c \approx 0.3116$~\cite{Stauffer1992}, regardless of the interval. Interestingly also, on the vertical axis [corresponding to $\varepsilon=0$ for the Hamiltonian of Eq.~\eqref{eq:sym_Hamiltonian}], the known value of the critical disorder $W_c \approx 16.5$ seems to correspond almost exactly to the point where $\Omega_{6+W/2}$ occupies the entire domain, i.e., to the value of $W$ for which $1/u \le 6 + W/2$ everywhere.

In conclusion, we have computed the percolation threshold of the LL-based effective potential $1/u$ in the 3D~Anderson tight-model, for uniform and binary disorders. In both cases, the similarity between this threshold and the ME on a large part of the phase diagram allows us to revisit our understanding of the Anderson transition. Even in the uniform case (or box disorder) where only one line separates the localized and delocalized phases, the ME in Fig.~\ref{fig:uniform_ME} consists in fact of two different parts corresponding to two different localization mechanisms: (i) a part close the edge of the spectrum (for values of the disorder strength such that $0 \le W \lesssim 14$), which can be interpreted from the LL~standpoint as the opening of a classically confining region, and (ii) a part of the line (for $14 \lesssim W \le 17$), specific to the discrete tight-binding model, where a disorder larger than the critical value blocks the existence of delocalized waves. These two parts are connected to form one single transition line in the Anderson uniform case, but are disjoint in the Anderson binary case. A large fraction of the ME (the one remaining in the continuous setting) is essentially of semi-classical nature, although one needs the renormalization of the disorder offered by the LL theory to observe it. This work opens the possibility to compute the critical exponents associated to the divergence of the size of $\Omega_E$ in the infinite domain near the percolation threshold, for any type of disorder or lattice, and to compare them with the known approximated values of the critical exponents of Anderson transition.

\section*{Acknowledgments}

M.F., P.P., and S.M. are supported by a grant from the Simons Foundation (1027116 Filoche; 601944, Filoche; 563916, Mayboroda). S.M. is supported by the NSF RAISE-TAQS grant DMS-1839077.

\appendix


\subsection*{Appendix A: Exhibiting $1/u$ as an effective potential}

In the discrete setting, the conjugation of the Hamiltonian~$\hat{H}$ by the localization landscape (LL)~$u$ is performed in the following way:
\begin{align}\label{eq:App_effective_1}
&(\hat{H}\psi)_n = -\sum_{m \in \ev{n}} (\psi_m - \psi_n) + V_n \psi_n \nonumber\\
&= -\sum_{m \in \ev{n}} (u_m \varphi_m - u_n \varphi_n) + V_n u_n \varphi_n \nonumber\\
&= -\sum_{m \in \ev{n}} \left( u_m ( \varphi_m - \varphi_n) + \varphi_n (u_m - u_n) \right) + V_n u_n \varphi_n \nonumber \\
&= -\sum_{m \in \ev{n}} u_m ( \varphi_m - \varphi_n) \nonumber\\
&\qquad \qquad + \varphi_n \left( -\sum_m (u_m - u_n) + V_n u_n \right) \nonumber\\
&= -\sum_{m \in \ev{n}} u_m ( \varphi_m - \varphi_n)  ~+~ \varphi_n \,.
\end{align}

Taking the inner product of this with the quantum state~$\psi$ leads to
\begin{align}\label{eq:App_effective_2}
\expval{\hat{H}}{\psi} = &\sum_n u_n \varphi_n \left( -\sum_{m \in \ev{n}} u_m ( \varphi_m - \varphi_n) + \varphi_n \right) \nonumber\\
= - &\sum_{\substack{n \\ m \in \ev{n}}} u_n u_m \varphi_n (\varphi_m - \varphi_n) + \sum_n u_n \varphi_n^2 \,.
\end{align}
The first term is a sum over all pairs of sites $(n,m)$ which are nearest neighbors. In fact, one can note that in this sum, each link $(n,m)$ appears twice. Grouping both terms relating to $(n,m)$ gives:
\begin{align}\label{eq:App_effective_3}
\expval{\hat{H}}{\psi} = \sum_{(m,n)} u_n u_m \left(\frac{\psi_n}{u_n} - \frac{\psi_m}{u_m} \right)^2 + \sum_n \frac{\psi_n^2}{u_n} \,.
\end{align}

\subsection*{Appendix B: Statistical symmetry of the spectrum}
In a continuous setting, the Hamiltonian is bounded only from below, which leads to the definition of one landscape. In a lattice, the presence of two band edges ($-6-W/2$ and $6+W/2$) permits to define two localization landscapes. One is obtained by shifting the Hamiltonian in order to push 0 out of the spectrum to the left. This is what is done in Eq.~\eqref{eq:asym_Hamiltonian}. Another landscape is obtained by pushing 0 to the right of the spectrum, which is done by shifting the sign of the Hamiltonian and adding $6 + W/2$ to it~\cite{Lyra2015EPL}. Indeed, the spectrum of the Hamiltonian $\hat{H}_0$ introduced in Eq.~\eqref{eq:sym_Hamiltonian} possesses a statistical symmetry around $E=0$ when the distribution of $\nu_n$ is even. To show that, one defines the transformation $T$ as
\begin{equation}
(T \psi)_n = (-1)^\abs{n} \, \psi_n \,,
\end{equation}
where $n=(n_1, n_2, \cdots, n_d)$ and $\abs{n} = n_1 + n_2 + \cdots n_d$. For each eigenstate~$\psi$ of $\hat{H}_0$ of eigenvalue~$E$, one can compute 
\begin{align}
&-\sum_{m \in \ev{n}} (T\psi)_m - W \nu_n \, (T\psi)_n \nonumber \\
=& -\sum_{m \in \ev{n}} (-1)^\abs{m} \psi_m - W \nu_n \, (-1)^\abs{n} \psi_n \nonumber\\
=& (-1)^\abs{n} \left[ \sum_{m \in \ev{n}} \psi_m - W \nu_n \, \psi_n\right] \nonumber \\
=& - (-1)^\abs{n} E \psi_n = -E \, (T\psi)_n \,.
\end{align}
One sees therefore that $T\psi$ is eigenfunction (with energy~$-E$) of a Hamiltonian similar to $\hat{H}_0$ with a random potential $-W \nu$ instead of $W \nu$. If $\nu$ is a random variable symmetrically distributed around $0$, then the aforementioned Hamiltonian is another realization of $\hat{H}_0$, hence the statistical symmetry of the spectrum. The second landscape can thus be defined as the solution to:
\begin{equation}
\left(-\hat{H}_0 + 6 + \frac{W}{2} \right) u  = 1
\end{equation}
The disordered potential for this second landscape is the opposite of the initial potential, and has exactly the same statistical properties due to the symmetry of the box distribution in $[-W/2,W/2]$. We can therefore use the results of one landscape (left side of the energy-disorder diagram) and plot it on the right side of the diagram. If the random law were not symmetrical by a sign change, the mobility edge would not be symmetrical in the energy-disorder diagram, and there would be two different landscapes with different statistical properties, each predicting one side of the mobility edge.

Due to the nature of the transformation~$T$, this observation can be in fact generalized to any lattice which is a bipartite graph.


\bibliography{Filoche_ME}

\end{document}